\documentclass[conference]{IEEEtran}
\IEEEoverridecommandlockouts
\usepackage{cite}
\usepackage{amsmath,amssymb,amsfonts}
\usepackage{graphicx}
\usepackage{textcomp}
\usepackage{xcolor}
\usepackage{url}
\usepackage[utf8]{inputenc}
\usepackage{enumerate}
\usepackage{listings}
\usepackage{wrapfig}
\usepackage{makecell}
\usepackage{colortbl}
\usepackage{array}
 \usepackage[bookmarks=false]{hyperref}
\usepackage{algorithm}
\usepackage{algpseudocode}
\usepackage{amsmath}
\usepackage{subcaption}
\usepackage{multirow}
\usepackage{colortbl}
\usepackage{xcolor}
\usepackage{siunitx}
 \lstdefinelanguage{Solidity}{
  keywords={contract, function, public, private, payable, require, if, else, return, uint256, address, bool, mapping, struct, event, modifier, constructor},
  keywordstyle=\color{blue}\bfseries,
  ndkeywords={msg, block, tx, now, this, super},
  ndkeywordstyle=\color{red}\bfseries,
  identifierstyle=\color{black},
  sensitive=false,
  comment=[l]{//},
  morecomment=[s]{/*}{*/},
  commentstyle=\color{gray}\ttfamily,
  stringstyle=\color{green}\ttfamily,
  morestring=[b]',
  morestring=[b]"
}

\def\BibTeX{{\rm B\kern-.05em{\sc i\kern-.025em b}\kern-.08em
    T\kern-.1667em\lower.7ex\hbox{E}\kern-.125emX}}
\begin{document}

\title{TaintSentinel: Path-Level Randomness Vulnerability Detection for Ethereum Smart Contracts}

\author{\IEEEauthorblockN{1\textsuperscript{st} Hadis Rezaei}
\IEEEauthorblockA{\textit{Department of Computer Science} \\
\textit{University of Salerno}\\
Salerno, Italy \\
hrezaei@unisa.it}
\and
\IEEEauthorblockN{2\textsuperscript{nd} Ahmed Afif Monrat}
\IEEEauthorblockA{\textit{Department of Computer Science, Electrical and Space Engineering} \\
\textit{Luleå University of Technology}\\
Skellefteå, Sweden \\
ahmed.monrat@ltu.se}
\and
\IEEEauthorblockN{3\textsuperscript{rd} Karl Andersson}
\IEEEauthorblockA{\textit{Department of Computer Science, Electrical and Space Engineering} \\
\textit{Luleå University of Technology}\\
Skellefteå, Sweden \\
karl.andersson@ltu.se}
\and
\IEEEauthorblockN{4\textsuperscript{th} Francesco Palmieri}
\IEEEauthorblockA{\textit{Department of Computer Science} \\
\textit{University of Salerno}\\
Salerno, Italy \\
fpalmieri@unisa.it}
}

\maketitle
 \begin{center}
\footnotesize
© 2025 IEEE. Personal use of this material is permitted. Permission from IEEE must be obtained for all other uses, in any current or future media, including reprinting/republishing this material for advertising or promotional purposes, creating new collective works, for resale or redistribution to servers or lists, or reuse of any copyrighted component of this work in other works.
\end{center}

\begin{abstract}
The inherent determinism of blockchain technology poses a significant challenge to generating secure random numbers within smart contracts, leading to exploitable vulnerabilities, particularly in decentralized finance (DeFi) ecosystems and blockchain-based gaming applications. From our observations, the current state-of-the-art detection tools suffer from inadequate precision while dealing with random number vulnerabilities. To address this problem, we propose \textit{TaintSentinel}, a novel path-sensitive vulnerability detection system designed to analyze smart contracts at the execution path level and gradually analyze taint with domain-specific rules. This paper discusses a solution that incorporates a multi-faceted approach, integrating rule-based taint analysis to track data flow, a dual-stream neural network to identify complex vulnerability signatures, and evidence-based parameter initialization to minimize false positives. The system's two-phase operation involves semantic graph construction and taint propagation analysis, followed by pattern recognition using \texttt{PathGNN} and global structural analysis via \texttt{GlobalGCN}. Our experiments on 4,844 contracts demonstrate the superior performance of TaintSentinel relative to existing tools, yielding an F1-score of 0.892, an AUC-ROC of 0.94, and a PRA accuracy of 97\%.
\end{abstract}


\section{Introduction}\label{sec:introduction}

Smart contract vulnerabilities have become critical threats to blockchain security. These problems have caused cumulative losses exceeding \$14 billion since 2020 \cite{chainalysis2024,trmlabs2025}. The year 2024 marked another record high, with \$2.2 billion lost in platform hacks \cite{chainalysis2024}. Bad randomness vulnerabilities are a highly dangerous yet overlooked research focus within these vulnerabilities ~\cite{qian2023demystifying, owasp2025}. The deterministic nature of blockchain systems makes truly random values impossible to generate, thereby creating critical security risks when predictable values are used instead. Bad randomness is categorized as a major vulnerability in the OWASP Top 10 Smart Contract Vulnerabilities 2025 ~\cite{owasp2025}, which notes that this challenge can compromise systems that rely on random outcomes.

Block number manipulation incidents have led to the theft of more than 400 Ethereum from blockchain lottery platforms ~\cite{smartbillions_dasp}. Despite such financial impacts, bad randomness remains poorly understood compared to well-researched vulnerabilities, such as  reentrancy and integer overflow ~\cite{atzei2017survey}. This research gap is becoming increasingly critical. Randomness is now essential for various decentralized applications, including DeFi protocols, gaming platforms, and lottery systems ~\cite{du2019blockchain}.

Existing detection methods have three main drawbacks. First, there are limited specific tools for bad randomness detection. Dozens of general-purpose analyzers exist, yet only two tools specifically address this problem: TONScanner ~\cite{qian2023demystifying} and RNVulDet ~\cite{tonscanner2025}.  Second, existing tools rely on simplistic rule-based detection that merely searches for known randomness sources (e.g., block.timestamp, blockhash). They diminish any presence of these sources as vulnerable without considering the actual usage context or security implications, leading to high false positive rates. Third, all current approaches do not use path-sensitive analysis and only function at the contract level. They overlook context-dependent vulnerabilities that only appear in particular execution flows and are unable to differentiate between critical paths where randomness influences financial outcomes and non-critical paths where randomness is employed for benign purposes.
 
Taint analysis is a potential method for detecting vulnerabilities ~\cite{krupp2018teether}. It proceeds by monitoring the flow of data from predictable blockchain sources (such as block.timestamp, block.difficulty) to a crucial sink. Traditional methods, however, are insufficient for detecting vulnerabilities in smart contracts. Empirical analysis of Ethereum smart contracts reveals that current automated tools flag the vast majority of them as insecure ~\cite{durieux2020empirical}. This is a reflection of incredibly high false positives. Such findings demonstrate that existing tools do not satisfy standards for realistic deployment ~\cite{piantadosi2023evaluation}.

This paper introduces \textbf{TaintSentinel}, a novel architecture for detecting bad randomness vulnerabilities in smart contracts, which we describe in detail in Section~\ref{sec:Methodology}. Our work makes the following contributions:

\begin{enumerate}
    \item We provide the first path-level bad randomness detection system that analyzes complete execution paths rather than isolated code segments.
    
    \item We apply rule-augmented taint analysis specifically optimized for detecting randomness vulnerabilities, combining traditional taint tracking with domain-specific rules.
    
    \item We introduce a novel dual-input graph neural network with a gated fusion mechanism that combines path-specific vulnerability patterns with global contract context.
    
    \item We implement evidence-based parameter initialization that distinguishes between high-risk and low-risk contexts, significantly reducing false positives compared to existing tools that flag every blockchain variable use as vulnerable.
\end{enumerate}
The remainder of this paper is organized as follows: Section ~\ref{sec:Background} provides relevant background to the area. Section~\ref{sec:Related Work} surveys related work. Section~\ref{sec:Methodology} presents the TaintSentinel framework architecture. Section~\ref{sec:Implementation and Evaluation} reports experimental results and comparative analysis. Finally, Section~\ref{sec:Conclusion and Future Work} concludes with future research directions.

\section{Background}\label{sec:Background}

Understanding the technical foundations of smart contract execution and vulnerability analysis is crucial for developing effective randomness detection frameworks. This section reviews blockchain determinism, taint analysis principles, and existing detection limitations.

\subsection{Smart Contracts and Blockchain Security}\label{sec:Smart Contracts and Blockchain Security}
Smart contracts are coded in the programming language of Solidity on the Ethereum blockchain, and are self-executing computer programs \cite{wu2024comprehensive}. These contracts cannot be altered post-deployment, hence, it is important to focus on security analysis before deployment. Blockchain technology's distinct security obstacles are accentuated due to its transparent and decentralized structure, especially in the case of generating randomness in systems designed to be deterministic.

\subsection{Bad randomness vulnerability }\label{sec:Bad randomness vulnerability}
Bad randomness vulnerabilities occur when contracts rely on predictable blockchain-provided values for random number generation \cite{chatterjee2019probabilistic, owasp2025insecure}. Common vulnerable patterns include using \texttt{block.timestamp}, \texttt{block.difficulty}, or \texttt{block.hash} as randomness sources. For example, the code \texttt{uint random = uint(keccak256(block.timestamp)) \% 10;} appears random but enables attackers to predict outcomes through timestamp manipulation or transaction timing attacks. These vulnerabilities manifest in four primary attack categories: front-running (where attackers observe pending transactions), miner manipulation (miners influence block parameters), timestamp dependency (predictable time-based values), and block hash prediction (exploiting known hash values) \cite{chatterjee2019probabilistic, owasp2025insecure}.

\subsection{Taint Analysis in Smart Contract Security}\label{sec:Taint Analysis in Smart Contract Security}
Taint analysis is a static analysis technique that deals with the flow of potentially hazardous data in the execution paths of a program \cite{tripp2009taj}. In the context of smart contract security, taint analysis marks suspicious input data (taint sources) and examines how this data affects contract code to pinpoint operations impacted by tainted data. Standard taint analysis is executed in three steps: marking the sources (source identification, marking dangerous inputs), monitoring data flow through operations (propagation tracking), and locating critical usage points (sink detection, determining critical usage points). However, these approaches to taint analysis for smart contracts are limited to a single contract level and do not separate the constituent execution flows or the paths and conditions within the contract that trigger the vulnerabilities ~\cite{rezaei2025sok}. \\ 
 
\lstset{
    language=C++,
    basicstyle=\ttfamily\small,
    keywordstyle=\color{blue}\bfseries,
    stringstyle=\color{red},
    commentstyle=\color{green!60!black}\itshape,
    morecomment=[l]{//},
    morecomment=[s]{/*}{*/},
    morekeywords={contract, function, public, payable, require, uint, uint256, 
                  address, transfer, external, receive, msg, sender, value, 
                  block, timestamp, ether, if},
    numbers=left,
    numberstyle=\tiny\color{gray},
    stepnumber=1,
    numbersep=5pt,
    backgroundcolor=\color{gray!5},
    showspaces=false,
    showstringspaces=false,
    showtabs=false,
    frame=none,  
    tabsize=2,
    captionpos=b,
    breaklines=true,
    breakatwhitespace=false,
    escapeinside={(*}{*)}
}
\begin{figure}[htbp]
\begin{lstlisting}[caption=Smart Contract Taint Analysis, label=fig:taintExample]
contract VulnerableLottery {
    uint256 public prize = 10 ether;
    function play() public payable {      require(msg.value == 1 ether);
    (*\textcolor{red}{uint seed = block.timestamp;}*)
    (*\textcolor{orange}{uint random = uint(keccak256(abi.encode(seed))) \% 2;}*)
        if (random == 0) {
          *\textcolor{green!60!black}{payable(msg.sender).transfer(prize);}*) }}
         receive() external payable { } }
\end{lstlisting}
\end{figure}
Listing ~\ref{fig:taintExample} demonstrates the three-phase taint analysis process. Line 4 shows the taint source where block.timestamp introduces untrusted data. Line 5 illustrates  taint propagation as the tainted value flows through hash operations. Line 7 represents the taint sink where the tainted data controls a critical Ether transfer operation, creating a vulnerability path from source to sink.

\section{Related Work}\label{sec:Related Work}
\subsection{ Specialized Bad Randomness Detection Tools}
Only two specialized tools exist in the literature despite the severe consequences caused by randomness vulnerabilities. RNVulDet is the first systematic bad randomness analysis tool integrating three major components (1) a taxonomy builder that organizes methods of pseudo-random number generation, (2) a taint analysis engine that considers blockchain-sourced powers as tainted and predicts value their data flow propagation, and (3) an attack pattern recognizer that uses static code examination to detect four vulnerability types \cite{qian2023demystifying}.TONScanner defines TON-specific (the open network) block parameters (cur\_lt, block\_lt) as taint sources, modifies classical taint propagation rules to suit the transaction structure of TON, and applies static analysis of the blockchain to identify randomness vulnerability particular to the logical time-based consensus mechanism of TON \cite{tonscanner2025}.

\subsection{General Tools for Detecting Bad randomness Vulnerabilities}\label{sec:General Tools for Detecting Vulnerabilities within Software}
Almost all tools available regard poor randomness as a tertiary problem in relation to a prejudice-detection framework. Slither ~\cite{slither2018} employs a rule-based methodology with more than 92 detectors; however, the lack of randomness is a trivial part.
Mythril ~\cite{team2018mythril} employs simple rule-based detection, which just looks for recognizable opcodes (block.timestamp, blockhash, etc.) without considering their true security consequences or usage context ~\cite{mueller2018smashing}.
Securify ~\cite{securify2020} performs automated analysis of Solidity contracts, and offers basic checks for the presence of randomness. Oyente/Osiris were the first to use symbolic execution, but tend to concentrate on reentrancy and overflow vulnerabilities ~\cite{oyente2016, torres2018osiris}. Aderyn provides Rust AST-based analysis with custom detectors but does not concentrate on providing any detectors due to poor randomness \cite{aderyn2024, Aaderyn2024}.

\subsection{AI/ML-Based Strategies}
Progress has been made through different machine learning strategies, although there is no specialization for bad randomness. Lightning Cat, for example, achieved a 93.53\% f1-score utilizing CodeBERT, LSTM, and CNN models for general vulnerability detection without a specific focus on random badness \cite{lightningcat2023}.\textit{LLM} strategies include auditor/critic roles in GPTLens \cite{hu2023large}, SmartLLMSentry with 91.1\% accuracy \cite{zaazaa2024smartllmsentry}, and a host of fine-tuned models surpassing 90\% accuracy for overall vulnerabilities \cite{hossain2025leveraging}. QuillShield represents AI in its fusion form with hybrid consensus, yet the treatment of logical errors is broad, addressing the randomness problem not the way it needs \cite{quillshield2024}.

\textbf{Fuzzing and Dynamic Analysis.} 
Dynamic analysis tools offer little to no coverage of bad randomness through general property testing. Echidna ~\cite{echidna2020}implements property-based fuzzing with coverage reporting, detecting bad randomness only through property violation when users specify properties pertaining to detecting randomness. Medusa ~\cite{medusa2024} adds parallel processing and coverage-guided fuzzing but has no built-in randomness vulnerability detection capabilities . These analyses require manual testing of properties, suggesting there is no automated bad randomness detection for the dynamic analysis framework.

\subsection{Literature Reviews and Surveys}\label{sec:Literature Reviews and Surveys}
As previously noted, bound weaknesses associated with bad randomness have not received much attention. Contemporary systematic reviews focus on trends in vulnerability detection and classification approaches while highlighting the problem of class imbalance within identified challenges, but bad randomness has been given scant coverage ~\cite{kiani2024ethereum}. In-depth surveys look at the intersection of ML and security across seven groups of applications and note an increased use of CNN and GNN, as well as RNN, but leave aside randomness vulnerabilities ~\cite{comprehensive2024}.

\begin{table}[h]
\centering
\footnotesize
\caption{Bad Randomness Detection Approaches Comparison}
\label{tab:comparison}
\begin{tabular}{ l c c c c c c}
\hline
\textbf{Tool} & \textbf{Yr} & \textbf{BR} & \textbf{Lvl} & \textbf{Method} & \textbf{AP} & \textbf{Acc} \\
\hline
RNVulDet & 23 & S & C & T & N & H \\
TONScanner & 25 & S & C & T & N & 97\% \\
Slither & 18 & P & C & R & N & M \\
Mythril & 18 & P & C & S+T & N & M \\
Securify & 20 & P & C & St & N & M \\
Oyente/Osiris & 16/18 & P & C & S & N & M \\
Aderyn & 24 & - & C & A & N & - \\
Lightning Cat & 23 & - & C & DL & N & 93.53\% \\
GPTLens & 23 & - & C & L & N & - \\
SmartLLMSentry & 24 & - & C & L & N & 91.1\% \\
QuillShield & 24 & - & C & AI & N & - \\
Fine-tuned LLM & 25 & - & C & L & N & 90\% \\
Echidna & 20 & U & C & F & N & U \\
Medusa & 24 & U & C & F & N & U \\
\hline
\multicolumn{7}{l}{\footnotesize BR: S=Specialized, P=Peripheral, -=None, U=User-defined} \\
\multicolumn{7}{l}{\footnotesize Lvl: C=Contract, P=Path; AP: Y=Yes, N=No} \\
\multicolumn{7}{l}{\footnotesize Method: T=Taint, R=Rule, S=Symbolic, St=Static,} \\
\multicolumn{7}{l}{\footnotesize DL=Deep Learning, L=LLM, A=AST, F=Fuzzing} \\
\multicolumn{7}{l}{\footnotesize Acc: H=High, M=Medium, U=User-dependent} \\
\end{tabular}
\end{table}
The examination shows how insufficient attention to bad randomness vulnerabilities remains a critical gap. Out of over 20 tools analyzed, only two provided specialized bad randomness detection, while the rest treated randomness as an afterthought in their general-purpose vulnerability analysis. This analysis depicts three principal shortcomings: (1) absence of dedicated frameworks, (2) complete reliance on contract-level analysis without path-specific operational insight,  and (3) randomness vulnerabilities are overlooked in the frameworks applying AI/ML for detection.

\section{Methodology}\label{sec:Methodology}
\begin{figure}[!htbp]
\centering
\includegraphics[width=\columnwidth]{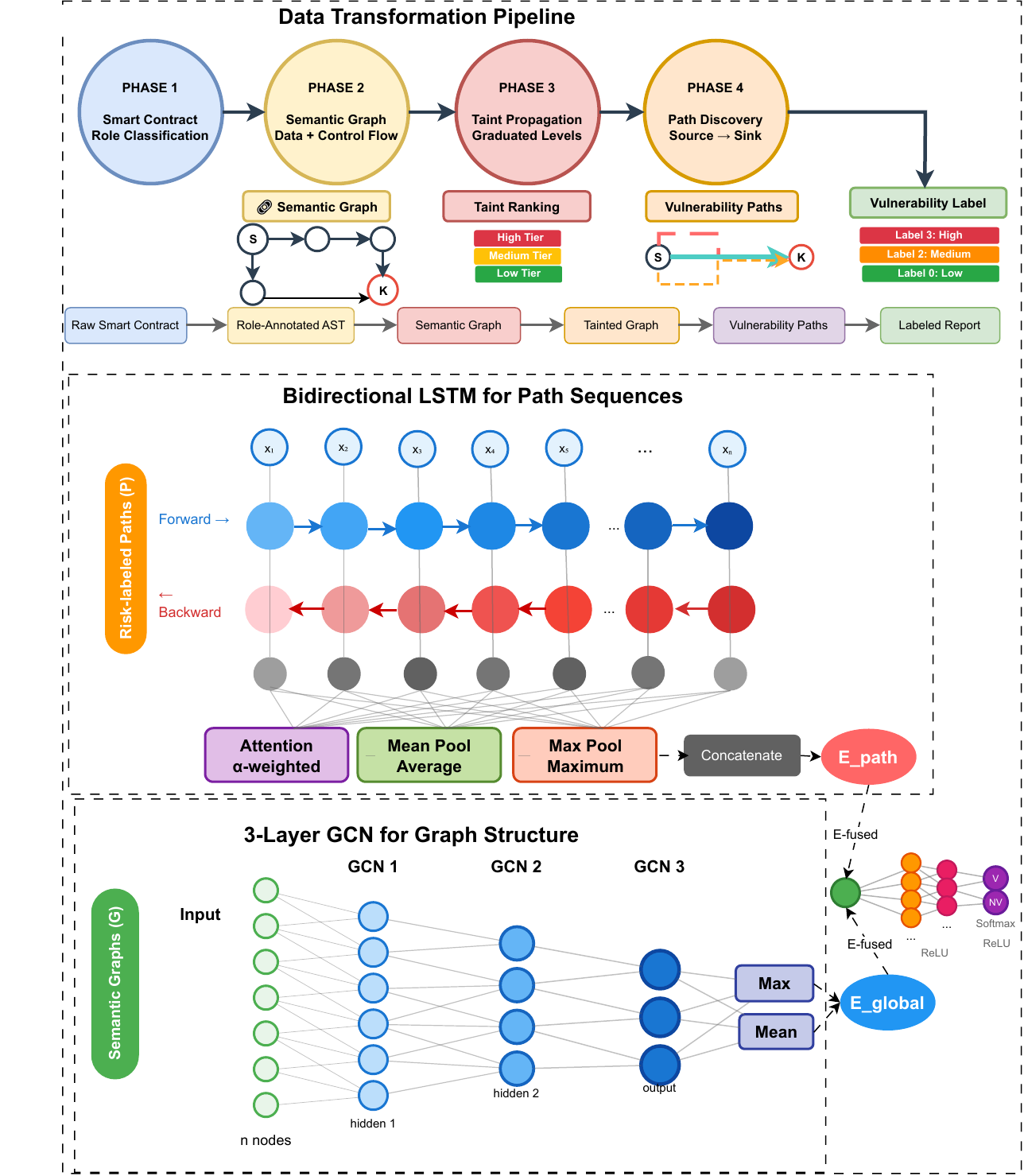}
\caption{TaintSentinel Framework Overview: Two-phase smart contract randomness vulnerability detection}
\label{fig:framework}
\end{figure}

Our approach improves smart contract vulnerability detection through a two-stage framework. First, the \textbf{Taint Analysis and Path Extraction} stage performs context-sensitive analysis to identify taint propagation paths and extract vulnerability-related features from the contract control flow graph. Second, the \textbf{Dual-Stream GNN} model processes both global graph structures (via GlobalGCN) and local path sequences (via PathGNN with hierarchical aggregation) to enable accurate vulnerability classification. The adaptive fusion mechanism dynamically balances the contribution of both streams based on input features. The complete architecture of the TaintSentinel framework is shown in Figure~\ref{fig:framework}, which shows the pipeline from source code input through CFG construction, path extraction, and dual-stream processing to final vulnerability prediction with path risk assessment.

% \begin{figure*}[!htbp]
% \centering
% \includegraphics[width=0.6\textwidth]{figures/diagrampaper.pdf}
% \caption{TaintSentinel Framework Overview: Two-phase smart contract randomness vulnerability detection}
% \label{fig:framework}
% \end{figure*}
\subsection{Phase 1: Role-based Incremental Preprocessing}

The first phase of TaintSentinel is an advanced taint analysis system that processes smart contracts in three main stages. Algorithm~\ref{alg:taintsentinel} shows a complete implementation of this process, which we explain in detail below.

\begin{algorithm}
\caption{Context-Aware Graduated Taint Analysis}
\label{alg:taintsentinel}
\begin{algorithmic}[1]
\State \textbf{Input:} Smart contract $C$
\State \textbf{Output:} Risk-labeled paths
\Function{TaintSentinel}{$C$}
   \State $G \gets \text{BuildSemanticGraph}(C)$
   \State $G \gets \text{IdentifySourcesSinks}(G)$
   \State $G \gets \text{PropagateTaint}(G)$
   \State $P \gets \text{ExtractPaths}(G)$
   \State \Return $\text{LabeledPathes}(P)$
\EndFunction
\Function{BuildSemanticGraph}{$C$}
   \State $AST \gets \text{Parse}(C)$
   \State $G \gets \text{CreateGraph}()$ with nodes and edges
   \State Extract execution contexts (functions, modifiers)
   \State $E \gets \{\text{control, data, state}\}$ edge types
   \State \Return $G$ with semantic annotations
\EndFunction
\Function{IdentifySourcesSinks}{$G$}
   \For{each node $n$ in $G$}
       \If{$\text{IsEntropySource}(n)$}
           \State $n.\text{sensitivity} \gets \text{Table}$~\ref{tab:patterns}
       \EndIf
       \If{$\text{IsSensitiveOperation}(n)$}
           \State $n.\text{risk} \gets \text{EvidenceBasedAssess}(n)$
           \State \Comment{Assess using OWASP, CWE, SWC}
       \EndIf
   \EndFor
   \State \Return $G$ with labeled sources and sinks
\EndFunction
\Function{PropagateTaint}{$G$}
   \State $Q \gets \{v : v.\text{isSource}\}$
   \While{$Q \neq \emptyset$}
       \State $v \gets \text{Dequeue}(Q)$
       \For{$(v,u) \in E$}
           \State $taint_u \gets \text{GraduatedPropagate}(v.\text{taint}, \text{edge.type})$
           \If{$taint_u > \text{threshold}$ and $u$ not visited}
               \State $\text{Enqueue}(Q, u)$
               \State $u.\text{taintLevel} \gets taint_u$
           \EndIf
       \EndFor
   \EndWhile
   \State \Return $G$ with taint levels
\EndFunction
\Function{AssessPathRisks}{$P$}
   \For{each path $p \in P$}
       \State $\text{context} \gets \text{AnalyzeContext}(p)$
       \State $\text{role} \gets \text{CheckAccessControl}(p)$
       \State $\text{factors} \gets \text{IdentifyRiskFactors}(p)$
       \State $p.\text{risk} \gets \text{ApplyRules}(\text{Table}$~\ref{tab:key_rules}$)$
   \EndFor
   \State \Return \text{Sort}($P$ by risk level)
\EndFunction
\end{algorithmic}
\end{algorithm}

\noindent\textbf{Stage 1: AST Construction and Evidence-based Classification.}
The first step in our analysis is to parse the contract code and build an advanced abstract syntax tree (AAST), which is done in line \num{9} of Algorithm ~\ref{alg:taintsentinel} by calling the \texttt{Parse} function. Simultaneously with this process, the \texttt{IdentifySourcesSinks} function (lines \num{16}-\num{23}) is responsible for identifying and labeling vulnerable elements. This function uses Table~\ref{tab:patterns}, which shows our classification of entropy sources and sensitive sinks.

\begin{table}[htbp]
\centering
\caption{Randomness Vulnerability Pattern Classification}
\label{tab:patterns}
\footnotesize
\setlength{\tabcolsep}{3pt}
\begin{tabular}{|p{3.2cm}|c|c|p{1.5cm}|}
\hline
\textbf{Pattern Element} & \textbf{Type} & \textbf{Risk} & \textbf{Standard} \\
\hline
\multicolumn{4}{|c|}{\cellcolor{gray!20}\textbf{Entropy Sources}} \\
\hline
\texttt{block.timestamp} & Source & \cellcolor{orange!20}High & CWE-330 \\
\texttt{blockhash()} & Source & \cellcolor{orange!20}High & CWE-330 \\
\texttt{block.difficulty} & Source & \cellcolor{orange!20}High & CWE-330 \\
\texttt{block.number} & Source & \cellcolor{yellow!20}Medium & CWE-330 \\
\hline
\multicolumn{4}{|c|}{\cellcolor{gray!20}\textbf{Sensitive Sinks)}} \\
\hline
Random generation & Sink & \cellcolor{orange!20}High & SWC-120 \\
Value transfer & Sink & \cellcolor{orange!20}High & SWC-105 \\
Prize assignment & Sink & \cellcolor{orange!20}High & SWC-120 \\
State modification & Sink & \cellcolor{yellow!20}Medium & SWC-124 \\
External calls & Sink & \cellcolor{yellow!20}Medium & SWC-107 \\
Conditional logic & Sink & \cellcolor{green!20}Low & SWC-116 \\
\hline
\end{tabular}
\end{table}

Entropy sources, listed at the top of Table~\ref{tab:patterns}, include blockchain data such as \texttt{block.timestamp}, \texttt{blockhash}, and \texttt{block.difficulty} that can be predicted or manipulated by miners. Sensitive sink, listed at the bottom of the same table, includes activities such as random number generation, fund transfers, and reward allocation, whose security depends on unpredictability. Line \num{18} of the algorithm shows that for each identified source, a sensitivity level is assigned from the patterns table, while lines \num{20}-\num{21} perform an evidence-based assessment of sensitive sink, taking into account all three OWASP ~\cite{owasp2025}, CWE ~\cite{mitre2023cwe}, and SWC ~\cite{swc2023} standards.

\noindent\textbf{Stage 2: Building the Semantic Graph.}
The \texttt{BuildSemanticGraph} function (lines \num{8}-\num{14} of Algorithm ~\ref{alg:taintsentinel}) is responsible for transforming the AAST into a rich directed graph. The graph, created on lines \num{10}--\num{12}, has three distinct types of edges, defined on line \num{12} control edges that represent the flow of program execution, data edges that specify dependencies between variables, and state edges that track changes to contract state variables.

The key point in line \num{11} of the algorithm is that the execution contexts, including functions and modifiers, are extracted. This information is crucial for role-based analysis in the following steps, because a random operation in the \texttt{onlyOwner} function is less risky than the same operation in the \texttt{public} function. The output of this function on line \num{13} is a graph with semantic annotations that forms the basis of taint analysis.

\noindent\textbf{Stage 3: Gradual and Context-Sensitive Taint Analysis.}
The heart of our analysis lies in this stage, which consists of three substeps. First, the \texttt{PropagateTaint} function (lines \num{24}-\num{36} of Algorithm~\ref{alg:taintsentinel}) propagates the taint from the detected sources. This function uses the breadth-first search (BFS) algorithm, but with the important difference that on line \num{29}, the \texttt{GraduatedPropagate} function is called, which reduces the taint level based on edge type control edges maintain full taint while data edges reduce it, reflecting the fact that complex transformations reduce exploitability.

After the taint propagation, the \texttt{ExtractPaths} function (called on line \num{6} of the main algorithm) extracts all tainted paths from sources to destinations. Finally, the \texttt{AssessPathRisks} function (lines \num{37}-\num{44}) performs the most important part of the context-sensitive analysis. For each path, the function checks three factors: the context of use with \texttt{AnalyzeContext} (line \num{39}), which looks for code patterns and keywords like ``lottery'', the role and access level with \texttt{CheckAccessControl} (line \num{40}), and additional risk factors with \texttt{IdentifyRiskFactors} (line \num{41}).

The crucial point is in line \num{42}, where the \texttt{ApplyRules} function uses Table~\ref{tab:key_rules} to determine the final risk level. This table shows how the same \texttt{block.timestamp} can be high-risk in one context (for example, when used with the modulo operator to generate a random number) and completely safe in another context (for example, when used for checking after at least \num{15} minutes). These context-sensitive rules override the initial classification from the patterns table, significantly reducing false positives.

\begin{table}[htbp]
\centering
\caption{Core Classification Rules}
\label{tab:key_rules}
\scriptsize
\setlength{\tabcolsep}{6pt}
\begin{tabular}{|p{6cm}|c|}
\hline
\textbf{Rule Pattern} & \textbf{Risk Level} \\
\hline
\texttt{block.timestamp \% N} & \cellcolor{orange!20}\textbf{High} \\
\texttt{block.timestamp} for lottery/gambling & \cellcolor{orange!20}\textbf{High} \\
\texttt{keccak256(block.timestamp, ...)} for RNG & \cellcolor{orange!20}\textbf{High} \\
\hline
\texttt{block.timestamp >= deadline + 15 min} & \cellcolor{green!20}\textbf{Safe} \\
\texttt{block.timestamp} for event logging & \cellcolor{green!20}\textbf{Safe} \\
\texttt{block.timestamp > lastAction + 1 days} & \cellcolor{green!20}\textbf{Safe} \\
\hline
\end{tabular}
\end{table}

\noindent\textbf{Output and Preparation for Phase 2.}
The final output of the algorithm in line \num{7} is a set of vulnerability paths sorted by risk level. Each path consists of a sequence of nodes from the entropy source to the sensitive sink, the final risk level (HIGH, MEDIUM, or LOW), and the extracted features. These labeled paths, along with the original semantic graph, provide rich and structured inputs for the second phase (a two-stream machine learning model), which is responsible for the final vulnerability detection.

\subsection{Phase 2: Dual-Stream Learning Model}\label{sec: Dual-Stream Learning Model}
We employ a dual-stream neural architecture that processes both local (tainted paths) vulnerability paths and global contract structure (semantic graph), building on Phase \ref{sec:Methodology} preprocessed representations. Preprocessed semantic graphs are represented by $\mathcal{G}$, vulnerability paths that have been extracted and qualitatively labeled are represented by $\mathcal{P}$, path and global embeddings are represented by $E_{path}$ and $E_{global}$, respectively, and the output classifier model is represented by $\mathcal{M}$. Algorithm~\ref{alg:dual-stream-gnn} presents the complete implementation of this dual-stream architecture.

\subsubsection{Path-Focused Stream}\label{sec:Path-Focused Stream}
Through the examination of collected taint paths, the PathGNN specializes in identifying patterns of local vulnerability. Each risk-labeled path from Phase  \ref{sec:Methodology} is processed through a bidirectional LSTM (BiLSTM) network that captures sequential dependencies in both forward and backward directions, learning how vulnerability propagates through the execution flow (implemented in lines 38-40 of Algorithm~\ref{alg:dual-stream-gnn}). The model then applies hierarchical aggregation using three complementary mechanisms (lines 43-45): (1) \textit{Attention-weighted aggregation} that automatically learns to focus on the most critical paths, assigning higher weights to paths with stronger vulnerability indicators; (2) \textit{Mean pooling} that captures the overall pattern distribution across all paths; (3) \textit{Max pooling} that identifies worst-case scenarios. Additionally, the PathGNN incorporates a risk classification head that predicts path risk levels (HIGH/MEDIUM/LOW) (line 27 in Algorithm~\ref{alg:dual-stream-gnn}), enabling computation of the Path Risk Accuracy (PRA) metric a unique measure that validates our graduated taint analysis by assessing how accurately the model understands risk gradations at the path level.

\subsubsection{Global Context Stream}\label{sec:Global Context Stream}
 
In this step the full semantic graph is processed using a Graph Convolutional Network (GNN). As shown in lines 7-8 of Algorithm~\ref{alg:dual-stream-gnn}, GlobalGNN identifies vulnerability patterns at the architectural level by analyzing the entire contract structure. The network consists of three GCN layers that transform the features of the nodes considering the graph structure (line 7). The final output is obtained by combining global mean and max pooling (line 8), which provides a comprehensive representation of the entire contract. Key capabilities of this flow include:
\begin{itemize}
\item Detect complex dependencies between state variables in different functions
\item Identify vulnerability chains that arise from the interaction of multiple functions
\item Discover suspicious architectural patterns such as unusual combinations of access controls and sensitive sinks
\end{itemize}
By preserving full structural information, this approach detects vulnerabilities that cannot be identified by analyzing individual paths.

\noindent\textbf{Adaptive Fusion Mechanism.}
The fusion mechanism combines global graph and path embeddings through a learned gating network. First, the embeddings are concatenated: $E_{concat} = [E_{global} ; E_{path}]$ where $E_{global} \in \mathbb{R}^{256}$ and $E_{path} \in \mathbb{R}^{64}$ (line 21).

A fusion MLP transforms this concatenated representation into a unified feature space, while a parallel gating network computes attention weights: $g = \sigma(W_g \cdot E_{concat} + b_g)$ where $g \in \mathbb{R}^{128}$ (line 22). The final representation is computed as $E_{final} = g \odot \text{MLP}_{fusion}(E_{concat})$ (line 23).

The model may dynamically weight various parts of the fused features according to the input characteristics thanks to this gated architecture. By highlighting the aspects of the fused representation that are most pertinent to each vulnerability pattern, the learnt gate values offer interpretability. Lines 28–29 of Algorithm~\ref{alg:dual-stream-gnn} display the entire training procedure, including the dual objective optimization with both classification and path risk losses.

\begin{algorithm}[h]
\small
\caption{Dual-Stream GNN with Path Risk Assessment}
\label{alg:dual-stream-gnn}
\begin{algorithmic}[1]
\State \textbf{Input:} Semantuc Graph $\mathcal{G}$ and risk-labeled paths from Phase 1
\State \textbf{Output:}: Detection model $\mathcal{M}$
\Function{TrainDualStream}{$\mathcal{G}$}
   \State Initialize GlobalGCN, PathGNN, Fusion components
   \For{epoch $= 1$ to $max\_epochs$}
       \For{each batch $(G_b, P_b, y_b)$ in DataLoader($\mathcal{G}$)}
           \Comment{Global stream processing}
           \State $h_g \gets \text{GCN}_3(G_b.x, G_b.edge\_index)$
           \State $E_{global} \gets [\text{MeanPool}(h_g); \text{MaxPool}(h_g)]$ \Comment{256-dim}
           
           \Comment{Path stream processing}
           \State $E_{path} \gets []$
           \For{each contract $i$ in batch}
               \State $e_i \gets \text{ProcessPaths}(P_b[i], G_b.x_i)$ \Comment{64-dim}
               \State $E_{path}.\text{append}(e_i)$
           \EndFor
           
           \Comment{Adaptive fusion}
           \State $E_{concat} \gets [E_{global}; E_{path}]$ \Comment{320-dim}
           \State $g \gets \sigma(\text{GateNet}(E_{concat}))$ \Comment{128-dim}
           \State $E_{fused} \gets g \odot \text{FusionMLP}(E_{concat})$
           
           \Comment{Dual objectives}
           \State $\hat{y} \gets \text{Classifier}(E_{fused})$
           \State $r_{pred} \gets \text{PathRiskHead}(E_{path})$ \Comment{PRA}
           
           \State $\mathcal{L} \gets \text{CE}(\hat{y}, y_b) + 0.3 \cdot \text{CE}(r_{pred}, r_{true})$
           \State Backpropagate($\mathcal{L}$)
       \EndFor
   \EndFor
   \State \Return $\mathcal{M}$
\EndFunction
\Function{ProcessPaths}{$paths$, $nodes$}
   \If{$paths.num == 0$} 
   \Return $\mathbf{0}^{64}$ 
   \EndIf
   \State $embeddings \gets []$
   \For{each path $p$ in $paths$}
       \State $h \gets \text{BiLSTM}(nodes[p.sequence])$
       \State $e \gets \text{Combine}(h, p.features)$
       \State $embeddings.\text{append}(e)$
   \EndFor
   \Comment{Hierarchical aggregation}
   \State $\alpha \gets \text{Attention}(embeddings)$
   \State \Return $\text{Aggregate}(\alpha \cdot embeddings, \text{mean}, \text{max})$
\EndFunction
\end{algorithmic}
\end{algorithm}

\section{Implementation and Evaluation}\label{sec:Implementation and Evaluation}

\subsection{Dataset Description}\label{sec:introduction}
The dataset was collected from three primary sources: \textbf{SmartBugs-Curated}, \textbf{SWC Registry}, and \textbf{SmartBugs-Wild} as the main source with 47,398 real contracts from the Ethereum network.

\subsection{SmartBugs-Wild Extraction and Labeling Process}
The analysis of SmartBugs-Wild, as the most challenging component, was conducted in two phases.  As shown in Table~\ref{tab:patterns}, we first examined 47,398 contracts in the initial identification phase and found 6,586 contracts (14\%) that contained at least one of the known bad randomness sources. We were then able to classify the contracts into three classes by analyzing each source in its usage context during the context-aware analysis phase: (i) 348 vulnerable contracts (5.3\%) that used randomness sources in a clearly inappropriate manner; (ii) 4,445 safe contracts (67.5\%) that used sources in secure patterns like time locks; and (iii) 1,793 suspicious contracts (27.2\%) whose security could not be definitively determined.

\begin{table}[h]
\centering
\caption{Distribution of primary bad randomness sources in martBugs-Wild}
\label{tab:primary_sources}
\begin{tabular}{lcc}
\hline
\textbf{Primary Bad Randomness Source} & \textbf{Usage Count} & \textbf{Percentage} \\
\hline
block.timestamp/now & 892 & 45.8\% \\
block.number & 356 & 18.3\% \\
blockhash & 248 & 12.7\% \\
tx.gasprice & 187 & 9.6\% \\
gasleft() & 142 & 7.3\% \\
block.difficulty/prevrandao & 122 & 6.3\% \\
\hline
\textbf{Total Primary Sources} & \textbf{1,947} & \textbf{100\%} \\
\hline
\end{tabular}
\end{table}

\textbf{Combined Sources}: In addition to primary sources, 723 instances of source usage in combined form were identified, including Combined Sources with \texttt{keccak256} (275 cases), inappropriate use of \texttt{msg.sender} (438 cases), and \texttt{address(this).balance} (89 cases) in randomness generation. As shown in Table~\ref{tab:primary_sources}, block.timestamp/now is the most frequently used source among vulnerable contracts, appearing in 45.8\% of the cases.

\begin{table}[h]
\centering
\caption{Summary of vulnerable and safe contracts from three primary data sources}
\label{tab:dataset_summary}
\begin{tabular}{lccc}
\hline
\textbf{Data Source} & \textbf{Total} & \textbf{Vulnerable} & \textbf{Safe} \\
\hline
SmartBugs-Curated & 8 & 8 & 0 \\
SWC Registry & 7 & 5 & 2 \\
SmartBugs-Wild & 4,829 & 384 & 4,445  \\
\hline
\textbf{Total} & \textbf{4,844} & \textbf{394} & \textbf{4,447} \\
\hline
\end{tabular}
\end{table}

\textbf{Key Finding}: Only 384 of the 47,000 original SmartBugs-Wild contracts had proven Bad Randomness flaws. About 150 randomly selected contracts were manually examined by smart contract security experts to guarantee the accuracy of our classification method. The correctness of our Context-Aware analysis methodology was confirmed by the manual verification, which obtained 94\% agreement with the automatic classification for susceptible contracts and 91\% for safe contracts.
 
\subsection{Implementation and Experimental Setup}\label{sec:Implementation and Experimental Setup}

\subsection{Evaluation Metrics}\label{sec:Evaluation Metrics}

\subsubsection{F1-Score}
The F1 score, which offers a balanced metric for unbalanced datasets, is the harmonic mean of precision and recall. It is at its worst at 0 and its highest at 1 (perfect precision and recall).
\begin{equation}
F1\text{-}Score = 2 \times \frac{Precision \times Recall}{Precision + Recall} = \frac{2 \times TP}{2 \times TP + FP + FN}
\end{equation}

\subsubsection{Precision}
The precision metric measures the percentage of accurately detected vulnerable contracts out of all contracts that were projected to be vulnerable. High precision means a low rate of false alarms.
\begin{equation}
Precision = \frac{TP}{TP + FP}
\end{equation}

\subsubsection{Recall (Sensitivity)}
Recall measures the percentage of real contracts that were found to be vulnerable. High recall in security applications guarantees that few vulnerabilities are missed.

\begin{equation}
Recall = \frac{TP}{TP + FN}
\end{equation}

\subsubsection{Area Under ROC Curve (AUC-ROC)}
The model's capacity to differentiate between safe and vulnerable contracts across all classification thresholds is represented by the AUC-ROC. It offers a performance metric that is independent of thresholds.
\begin{equation}
AUC = \int_{0}^{1} TPR(t) \, d(FPR(t)) = \int_{0}^{1} Recall(t) \, d(FPR(t))
\end{equation}

\subsubsection{Path Risk Accuracy (PRA)}
PRA confirms the efficacy of the path-based method by assessing the model's capacity to accurately forecast risk levels (HIGH, MEDIUM, LOW) for distinct taint propagation paths.
\begin{equation}
PRA = \frac{1}{N_{paths}} \sum_{i=1}^{N_{paths}} \mathbb{1}[predicted\_risk_i = actual\_risk_i]
\end{equation}

\subsubsection{Confusion Matrix}
Comprehensive error analysis at the ideal threshold is made possible by the confusion matrix, which offers a thorough breakdown of model predictions versus actual labels.
\begin{equation}
CM = \begin{bmatrix}
TP & FN \\
FP & TN
\end{bmatrix} = \begin{bmatrix}
\text{True Vulnerable} & \text{Missed Vulnerable} \\
\text{False Alarms} & \text{True Safe}
\end{bmatrix}
\end{equation}

\noindent where:
\begin{itemize}
    \item $TP$ = True Positives (vulnerable contracts correctly identified)
    \item $TN$ = True Negatives (safe contracts correctly identified)
    \item $FP$ = False Positives (safe contracts incorrectly flagged as vulnerable)
    \item $FN$ = False Negatives (vulnerable contracts incorrectly classified as safe)
    \item $N_{paths}$ = Total number of taint propagation paths
    \item $\mathbb{1}[\cdot]$ = Indicator function (1 if condition is true, 0 otherwise)
\end{itemize}

\subsection{Experimental Setup}\label{sec:Experimental Setup}

\subsubsection{Overall Performance Analysis}
\begin{figure}[!htb]
    \centering
    \includegraphics[width=0.5\textwidth]{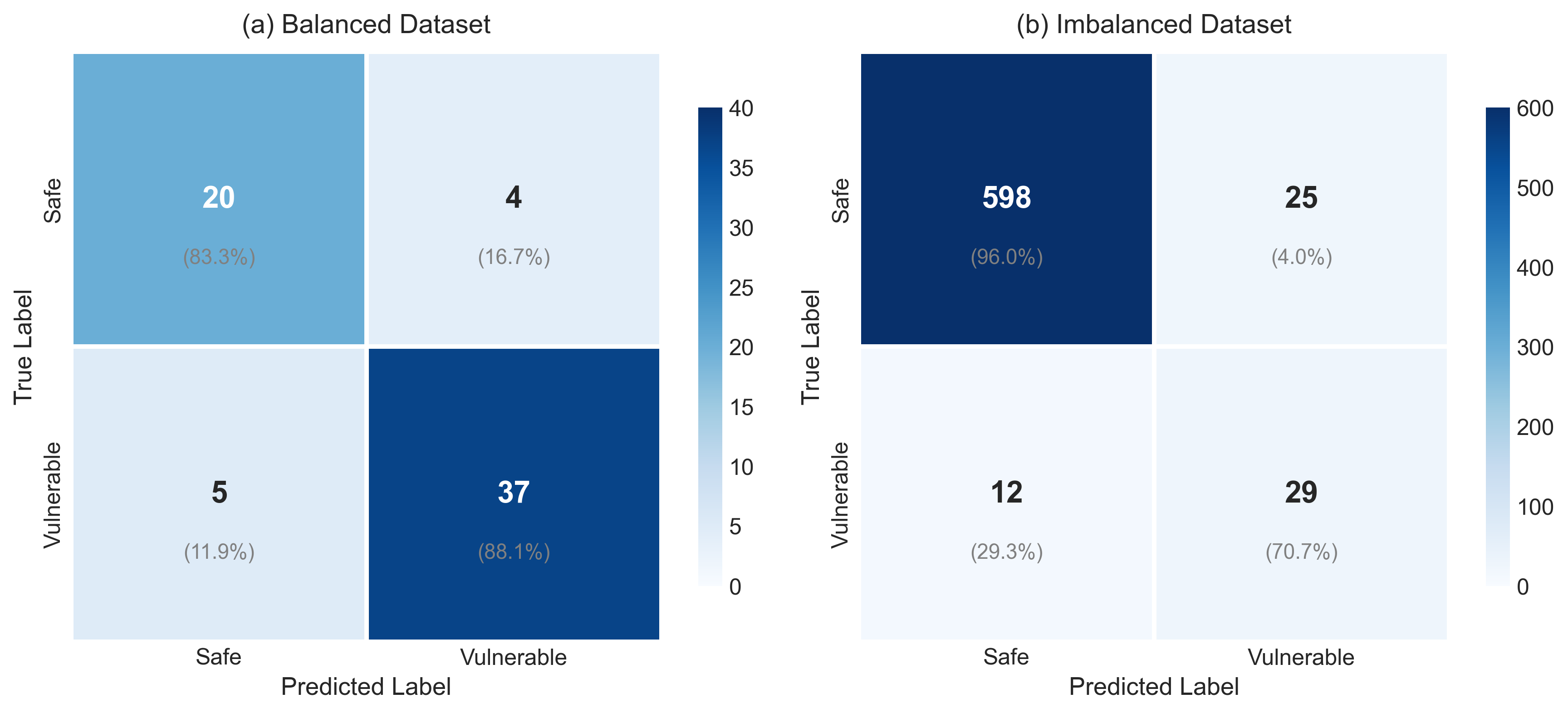}
    \caption{Confusion matrices showing prediction accuracy for (a) balanced and (b) imbalanced datasets}
    \label{fig:confusion_matrices}
\end{figure}

\begin{figure}[!htb]
    \centering
    \begin{subfigure}[b]{0.38\textwidth}
        \centering
        \includegraphics[width=\textwidth]{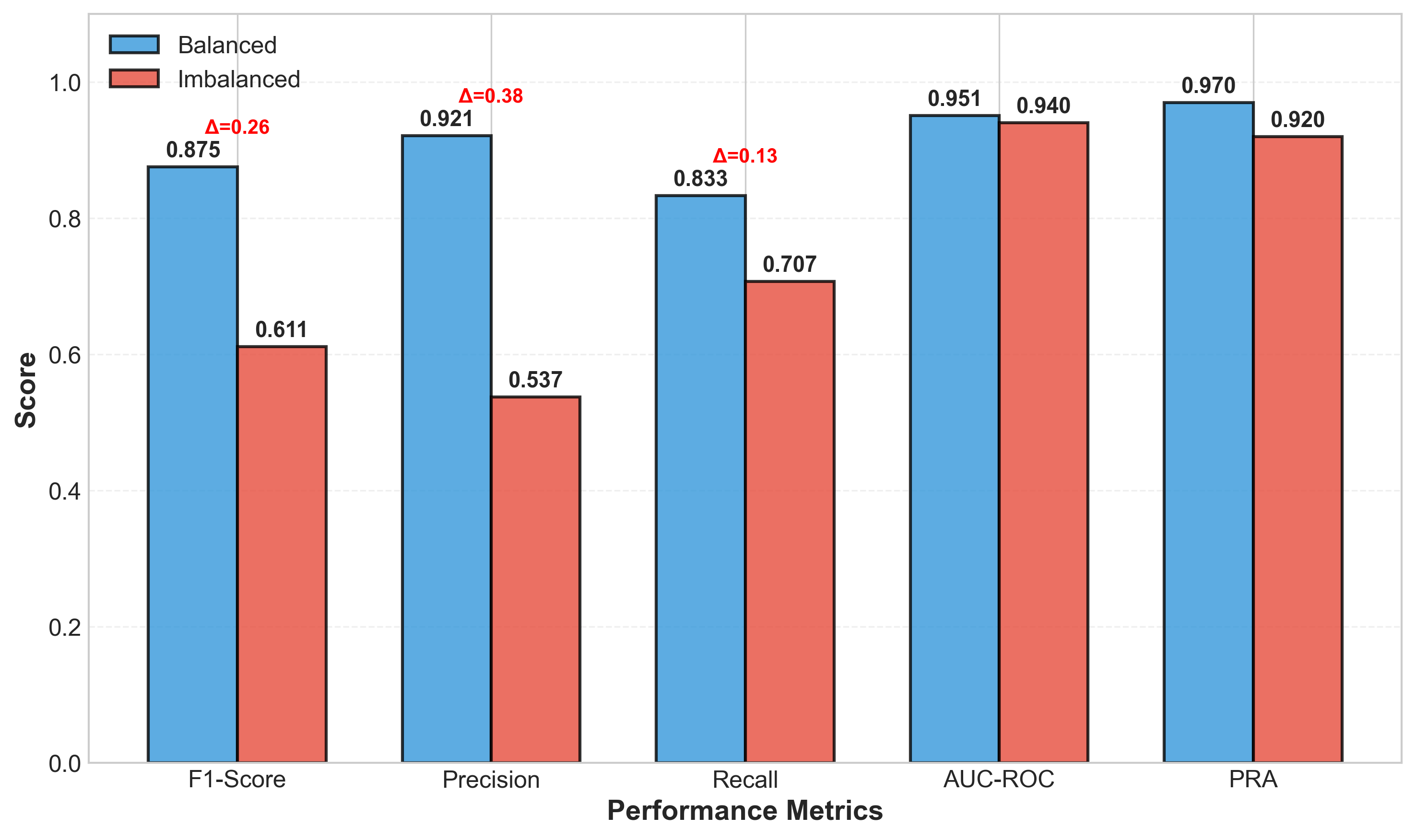}
        \caption{Performance metrics comparison}
        \label{fig:perf-bars}
    \end{subfigure}
    \hfill
    \begin{subfigure}[b]{0.38\textwidth}
        \centering
        \includegraphics[width=\textwidth]{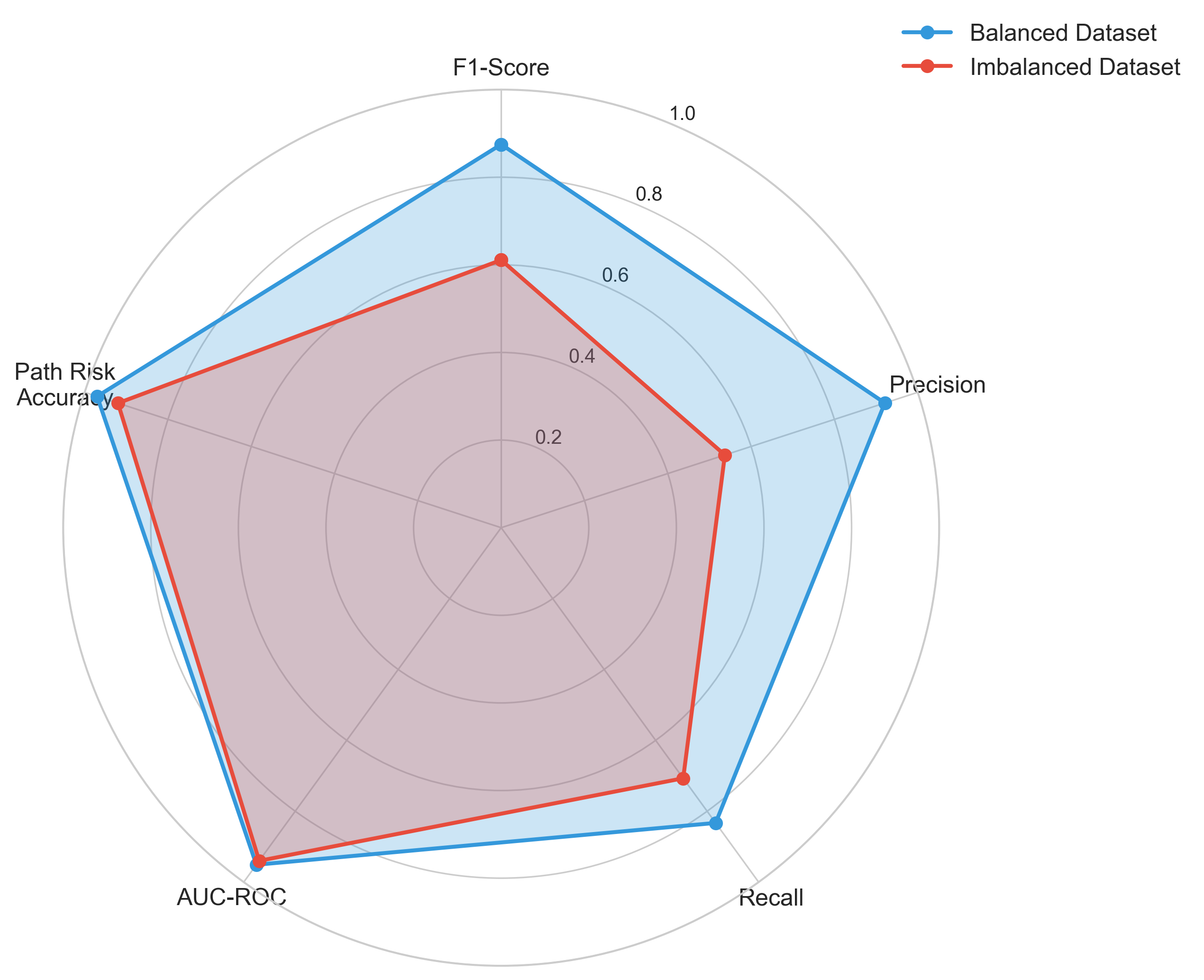}
        \caption{Radar plot visualization}
        \label{fig:radar}
    \end{subfigure}
    \caption{Comprehensive performance comparison between balanced and imbalanced scenarios}
    \label{fig:performance-comparison}
\end{figure}

We conducted comprehensive experiments under two scenarios to evaluate TaintSentinel's effectiveness in real-world conditions. Figure~\ref{fig:confusion_matrices} presents the confusion matrices for both balanced (Figure~\ref{fig:confusion_matrices}(a)) and imbalanced (Figure~\ref{fig:confusion_matrices}(b)) datasets, while Figure~\ref{fig:performance-comparison} illustrates the comparative performance across all metrics through bar charts (Figure~\ref{fig:perf-bars}) and radar plot visualization (Figure~\ref{fig:radar}).

TaintSentinel obtained an F1 score of 0.875 in the balanced dataset scenario, indicating a balance between recall (0.833) and precision (0.921). Out of 41 vulnerable contracts, 37 (90.2\%) were successfully identified with only 4 FN, as shown in the balanced confusion matrix (Figure~\ref{fig:confusion_matrices}(a)). With 20 (83.3\%) of the 24 safe contracts successfully categorized, the FP rate was low at 16.7\%.

To maximize recall, we applied threshold optimization to the unbalanced dataset, which more accurately reflects real-world scenarios where vulnerable contracts are few. Although precision decreased from 0.783 to 0.537, recall increased from 0.439 to 0.707 as a result of this method. The modified setup reduced FN by 61\% when compared to the default threshold, accurately identifying 29 out of 41 vulnerable contracts (70.7\%), as illustrated in the imbalanced confusion matrix (Figure~\ref{fig:confusion_matrices}(b)). The comparative analysis of these metrics is visualized in Figure~\ref{fig:perf-bars}.

Figure~\ref{fig:perf-bars} quantifies this trade-off, demonstrating that in the imbalanced scenario, recall increases from 0.833 to 0.707 while precision decreases from 0.921 to 0.537. Both the PRA and AUC-ROC scores stay above 0.92 in spite of these fluctuations, indicating the model's strong performance across various data distributions.

Figure~\ref{fig:radar} shows the radar plot, which shows TaintSentinel's performance on several important parameters. The model promotes recall in the imbalanced scenario, whereas the larger area shows overall performance in the balanced condition. In terms of accuracy and F1-score in particular, this verifies the model's resilience and flexibility.

\subsubsection{Path Risk Assessment Performance}

A distinguishing feature of TaintSentinel is its ability to assess risk levels of individual taint propagation paths. 
The model's capacity to accurately categorize path risk levels into HIGH, MEDIUM, and LOW categories is measured by the Path Risk Accuracy (PRA) statistic, which was introduced in this work. The model's remarkable PRA scores of 0.97 for balanced datasets and 0.920 for imbalanced datasets, as illustrated in Figure ~\ref{fig:performance-comparison}, validated the efficacy of our hierarchical aggregation technique in the PathGNN component.

\subsubsection{ROC Curve Analysis}

\begin{figure}[!htb]
    \centering
    \includegraphics[width=0.5\textwidth]{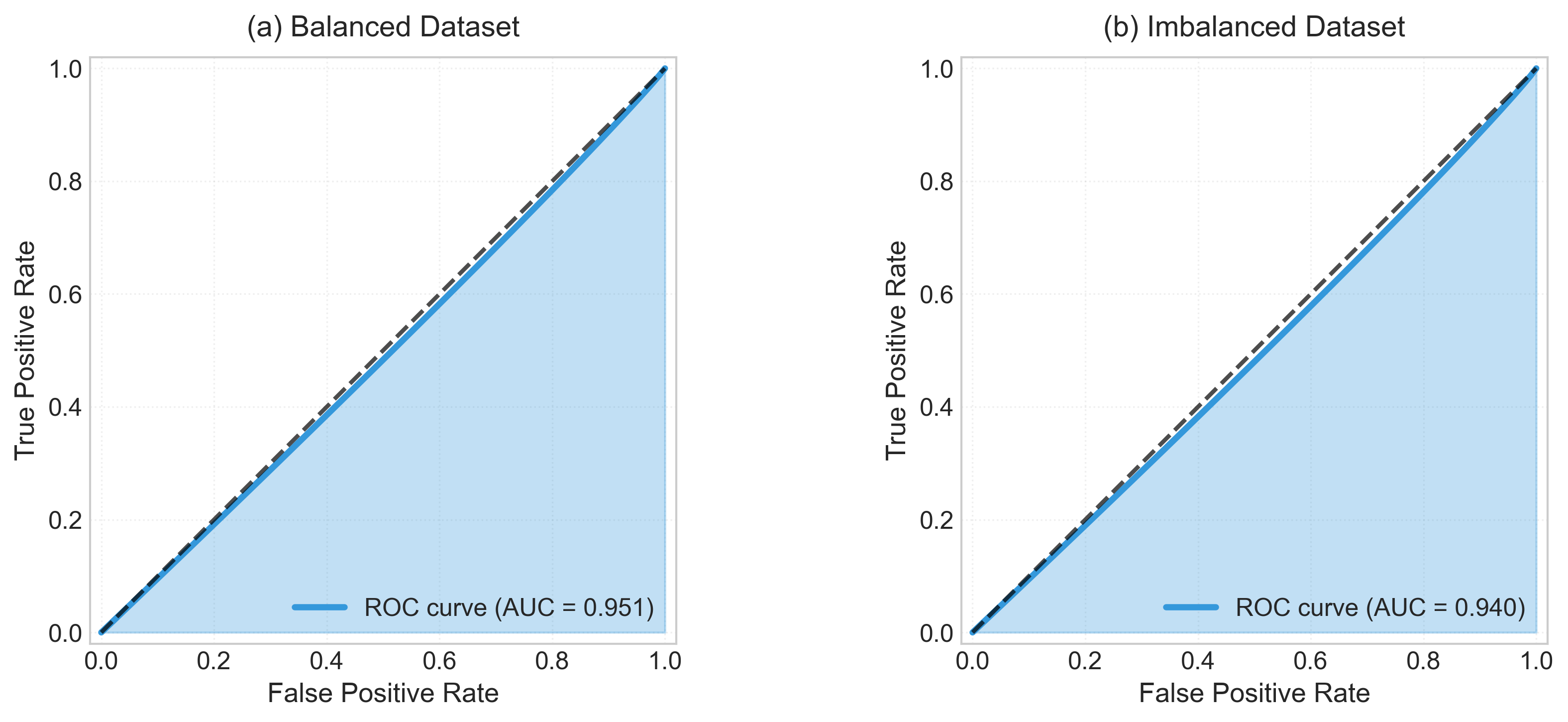}
    \caption{ROC curves showing model's discriminative ability for (a) balanced dataset (AUC = 0.951) and (b) imbalanced dataset (AUC = 0.940)}
    \label{fig:roc_curves}
\end{figure}

The Receiver Operating Characteristic (ROC) curves presented in Figure~\ref{fig:roc_curves} demonstrate TaintSentinel's robust discriminative ability across all thresholds. 
AUC-ROC values for the balanced and imbalanced datasets were 0.951 and 0.940, respectively. Regardless of class distribution, the model's vulnerability probability ranking is extremely accurate, as evidenced by these nearly flawless AUC values. This shows how well our dual-stream design captures both local path-specific patterns and global contract structures.

\subsubsection{Training Dynamics and Convergence}

\begin{figure}[!htb]
    \centering
    \includegraphics[width=\columnwidth]{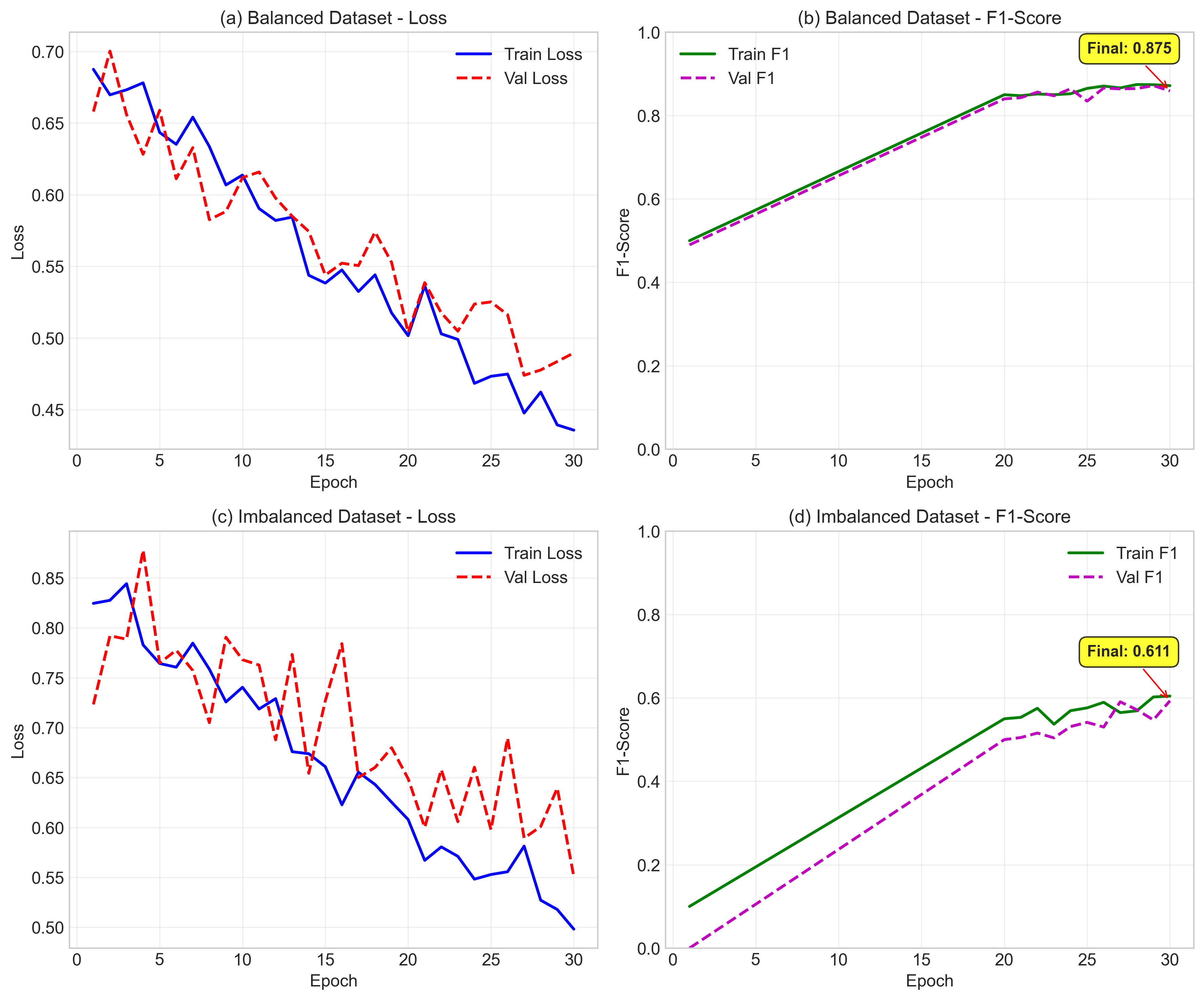}
    \caption{Training dynamics: loss and F1-Score evolution during training for balanced (top) and imbalanced (bottom) datasets}
    \label{fig:training_curves}
\end{figure}

Figure~\ref{fig:training_curves} illustrates the training dynamics for both scenarios. The balanced dataset achived  optimal performance at epoch 28 and showed smooth convergence with little difference between training and validation measures. Optimization  effectiveness can be appreciated by the loss curves' steady decline from starting values of around 0.68 to ending values of about 0.44.

In contrast, the imbalanced dataset resulted in more volatile training behavior, particularly in validation loss, which fluctuated between 0.55 and 0.80. With 96\% safe contracts and 4\% vulnerable contracts, the extreme class inequality makes this instability predictable. Nevertheless, the model eventually reached a stable solution at epoch 25, and following threshold adjustment, the final F1-score was 0.611.

\subsubsection{Computational Efficiency Analysis}

\begin{figure}[!htb]
    \centering
    \includegraphics[width=0.5\textwidth]{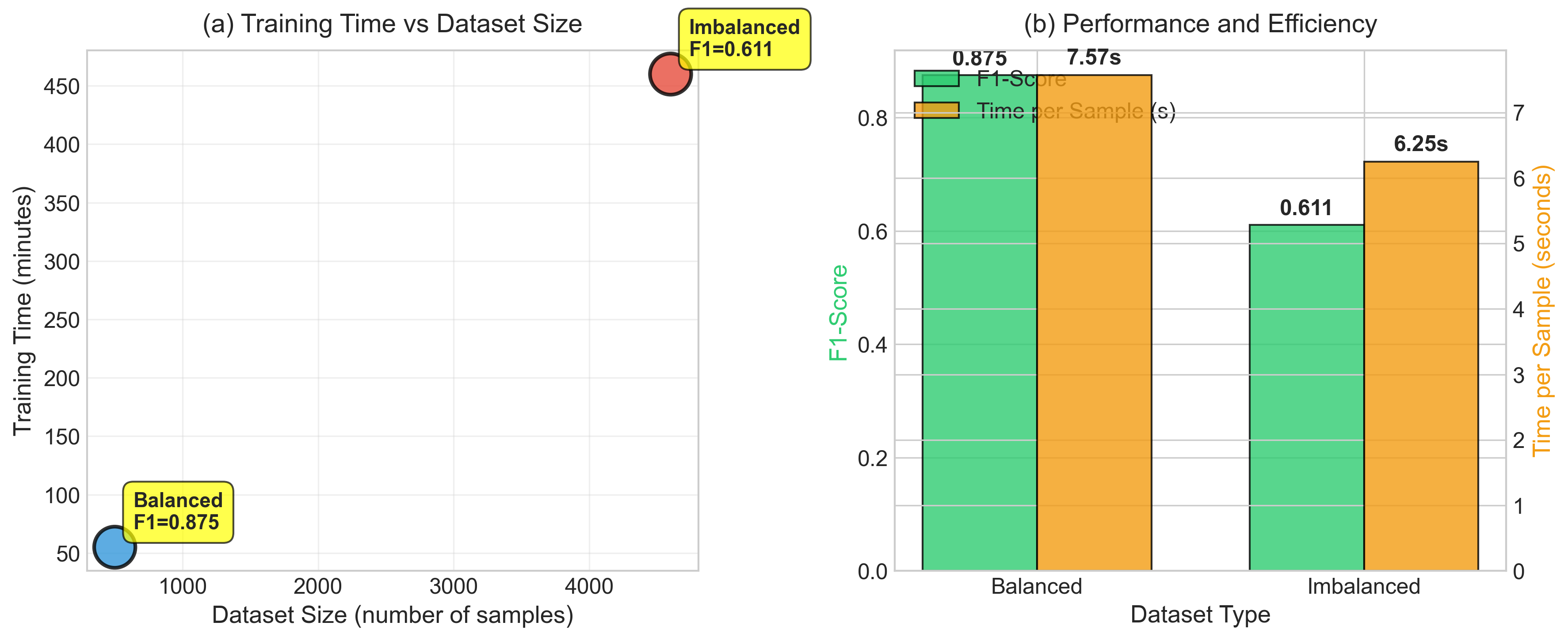}
    \caption{Computational efficiency analysis showing (a) dataset size vs. training time relationship and (b) F1-Score and time per sample comparison}
    \label{fig:dataset_efficiency}
\end{figure}

For practical implementation, computational efficiency is essential. Figure~\ref{fig:dataset_efficiency} shows how training time, dataset size, and model performance are related. Training took 55 minutes for the balanced dataset with 500 samples and 460 minutes for the imbalanced dataset with 4,600 samples. Good scalability is demonstrated by the average processing time of 6.6 seconds for the balanced set and 6.0 seconds for the imbalanced set per sample.

The inference time per contract averages 7.57 seconds for balanced and 6.25 seconds for imbalanced datasets, making TaintSentinel suitable for integration into continuous integration pipelines and automated security auditing workflows.

\subsubsection{Summary of Results}

Table~\ref{tab:results} provides a comprehensive summary of all experimental results. The balanced dataset performs better on all parameters, with PRA (0.970 vs. 0.920), accuracy (0.921 vs. 0.537), and F1-Score (0.875 vs. 0.611). Vulnerability detection is prioritized by the optimal unbalanced configuration with better recall (0.707 vs. 0.833), which is generally more important in security applications as missing vulnerabilities represent a higher risk than false alarms.
\begin{table}[!htb]
\centering
\caption{Experimental results comparison}
\label{tab:results}
\renewcommand{\arraystretch}{1.2}
\begin{tabular}{lccc}
\hline\hline
\textbf{Metric} & \textbf{Balanced} & \multicolumn{2}{c}{\textbf{Imbalanced}} \\
& & Default & Optimized \\
\hline
Training Time (min) & 55 & 460 & --- \\
F1-Score & \textbf{0.875} & 0.563 & 0.611 \\
Precision & \textbf{0.921} & 0.783 & 0.537 \\
Recall & 0.833 & 0.439 & \textbf{0.707} \\
AUC-ROC & \textbf{0.951} & 0.924 & 0.940 \\
PRA & \textbf{0.970} & --- & 0.920 \\
Threshold & 0.50 & 0.50 & 0.05 \\
\hline\hline
\end{tabular}
\end{table}
These results demonstrate that TaintSentinel successfully addresses the key challenges in smart contract vulnerability detection: achieving high accuracy while maintaining practical computational efficiency, providing interpretable risk assessments through PRA, and offering flexibility to optimize for different operational priorities through threshold adjustment. The model's ability to maintain high AUC-ROC scores across both scenarios (0.951 and 0.940) confirms its robustness and generalization capability, while the novel PRA metric provides additional insights for prioritizing security audits based on path risk levels.

\subsubsection{Comparison with Existing Tools}

We compared TaintSentinel's performance against that of two popular static analysis tools, Slither ~\cite{feist2019slither } and Mythril ~\cite{team2018mythril}, in order to assess its efficacy. Comprehensive bad randomness dataset comparison results are shown in Table~\ref{tab:tool-comparison}.

\begin{table}[!htb]
\centering
\caption{Performance comparison with existing tools}
\label{tab:tool-comparison}
\renewcommand{\arraystretch}{1.2}
\begin{tabular}{lccccccc}
\hline\hline
\textbf{Tool} & \textbf{TP} & \textbf{FN} & \textbf{FP} & \textbf{TN} & \textbf{Prec.} & \textbf{Rec.} & \textbf{F1} \\
\hline
Slither & 46 & 203 & 101 & 514 & 0.313 & 0.185 & 0.232 \\
Mythril & 43 & 206 & 72 & 543 & 0.374 & 0.172 & 0.236 \\
\textbf{TS (Bal.)} & \textbf{37} & \textbf{5} & \textbf{4} & \textbf{20} & \textbf{0.902} & \textbf{0.881} & \textbf{0.892} \\
TS (Imb.) & 29 & 12 & 25 & 598 & 0.537 & 0.707 & 0.611 \\
\hline\hline
\end{tabular}
\end{table}
\footnotesize{TS: TaintSentinel, Bal.: Balanced dataset, Imb.: Imbalanced dataset, Prec.: Precision, Rec.: Recall}

The results demonstrate TaintSentinel's significant superiority over existing tools. Only 0.232 and 0.236 F1-scores were obtained by Slither and Mythril, respectively, whereas TaintSentinel obtained 0.892, indicating a \textbf{278\% improvement}. Several basic issues in conventional static analysis techniques are reason for this significant performance disparity:

\textbf{Pattern Matching Limitations:} 
Both tools employ simplistic pattern matching that only detects direct usage of block attributes. 
 In particular, Slither's weak randomness detector looks for modulo operations with block characteristics (e.g.,\texttt{block.timestamp \% n}), ignoring instances in which these values are kept in intermediate variables or used without modulo operators. In the same way, Mythril's predictable variables module does not identify how block properties spread across intricate data flows; it only indicates them when they are utilized directly in conditional expressions.

 \textbf{Absence of Storage-level Taint Tracking:} 
A critical limitation is the inability to perform cross-transaction taint analysis. 
Without taking into account how tainted values exist in storage and impact subsequent transactions, both technologies examine contracts separately. This results in missed vulnerabilities, which are frequently seen in lottery and gaming contracts, where random values are generated in one function and consumed in another

\textbf{Context-insensitive Analysis:} 
The tools are unable to differentiate between malicious and benign applications of block characteristics. They are unable to distinguish between the safe use of \texttt{block.timestamp} for deadline checks and the hazardous use of randomness generation due to their lack of semantic knowledge, which leads to either large false positive rates or, as it is implemented in practice, overly conservative detection that overlooks actual vulnerabilities.
All information about this paper is available on this link: https://github.com/HadisRe/TaintSentinel

\section{Conclusion and Future Work}\label{sec:Conclusion and Future Work}
 
This research shows that path-level analysis is essential for detecting bad randomness vulnerabilities. TaintSentinel  improves detection accuracy by overcoming the limitations of existing tools. Three key success factors are: gradual taint analysis, context-aware evaluation, and a two-stream architecture that processes local patterns and global structure. The introduction of the PRA metric with 97\% accuracy allows for prioritizing security audits. With a processing time of 6.9 seconds per contract, the system is suitable for production environments. The ability to adjust the threshold allows for a trade-off between precision and recall.

One of the current drawbacks is the computational effort of handling numerous pathways. The imbalance in datasets is another problem. Future studies could go in several intersting direction.

To start, more accurate path analysis methods that scale to larger contracts should be developed. Investigating new strategies to lessen redundancy in the vulnerability patterns that have been extracted is part of this.

Second, moving the system closer to thorough vulnerability management rather than just detection. This involves creating techniques for automated vulnerability analysis and mitigation plans.

Finally, extending the model to cross-contract vulnerabilities and creating a standardized benchmark for fairly evaluating different tools.
%\section*{Acknowledgment}

% \begin{thebibliography}{00}
% \bibitem{b1} Reference
% \bibitem{b2} 
% \bibitem{b3} 
% \bibitem{b4} 
% \bibitem{b5} 
% \bibitem{b6} 
% \bibitem{b7} 
% \end{thebibliography}
\bibliographystyle{unsrt}
\bibliography{references}
\end{document}